\documentstyle{article}

         \def\be{\begin{equation}}
         \def\bea{\begin{eqnarray}}
         \def\ee{\end{equation}}
         \def\eea{\end{eqnarray}}

\begin{document}
\begin{titlepage}
\vspace*{5mm}
\begin{center} {\Large \bf On the phase structure of two--dimensional\\
\vskip 0.5cm
generalized Yang--Mills theories}
\vskip 1cm
{\bf Masoud Alimohammadi$^a$ \footnote {e-mail:alimohmd@theory.ipm.ac.ir} \&
Mohammad Khorrami$^b$ \footnote {e-mail:mamwad@iasbs.ac.ir}} \\
\vskip 1cm
{$^a$  Physics Department, University of Teheran, North Karegar Ave.,} \\
{Tehran, Iran }\\
{$^b$ Institute for Advanced Studies in Basic Sciences,}\\
{P. O. Box 159, Zanjan 45195, Iran.}\\
\end{center}
\vskip 2cm
\begin{abstract}
\noindent The phase structure of the generalized Yang--Mills theories is
studied, and it is shown that {\it almost} always, it is of the third order.
As a specific example, it is shown that all of the models with the
interaction $\sum_j (n_j-j+N)^{2k}$ exhibit third order phase transition.
($n_j$ is the length of the $j$-th row of the Yang tableau corresponding to
U($N$).) The special cases where the transition is not of the third order
are also considered and, as a specific example, it is shown that the model
$\sum_j (n_j-j+N)^2+g\sum_j (n_j-j+N)^{4}$ exhibits a third order phase
transition, except for $g=27\pi^2/256$, where the order of the transition is
$5/2$.
\end{abstract}

\begin{center}
\end{center}
\end{titlepage}
\newpage
\section{ Introduction }

Recent works by many authors suggest that the pure Yang--Mills theory in two
dimensions is equivalent to a string theory [1--7]. The reason to
investigate such a simple theory in such great detail is partly to shed
light on the strong--coupling limit of pure QCD in four dimensions. The
string picture of the two--dimensional Yang--Mills (YM$_2$) is also
interesting on its own, as an example for nonperturbative analysis of a
quantum field theory.

The starting point for making the correspondence of YM$_2$ with string
theory is the study the large--$N$ limit of YM$_2$. For example, as it is
shown in \cite{1}, \cite{3}, and \cite{4}, a gauge theory based on SU($N$)
is split at large $N$ into two copies of a chiral theory, which encapsulate
the geometry of the string maps. The chiral theory associated to the
Yang--Mills theory on a two--manifold $\cal M$ is a summation over maps from
the two--dimensional world sheet (of arbitrary genus) to the manifold
$\cal M$. This leads to a $1/N$ expansion for the partition function and
observables that is convergent for all of the values of area$\times$coupling
constant on the target space $\cal M$, if the genus is one or greater.

It is well known that YM$_2$ is defined by the Lagrangian tr$(F^2/4)$
on a Riemann surface, where $F$ is the field--strength tensor. This theory
is equivalent to the so--called $B$--$F$ theory, characterized by the
Lagrangian $i$tr$(BF)+$tr($B^2)$, integrated over the auxiliary field $B$.
This theory is essentially characterized by two important properties: the
lack of propagating degrees of freedom, and invariance under
area--preserving diffeomorphisms. This theory, however, is not unique in
possessing these two characteristics. These properties are also shared by a
large class of theories, called the generalized two--dimensional Yang--Mills
(gYM$_2$) theories. These theories are defined by replacing the term
tr($B^2$) in the YM$_2$ Lagrangian by an arbitrary class function of $B$
\cite{8}.

Many of the physical features of gYM$_2$ have been studied. Several aspects
of the partition function of these theories has been discussed in [8--10].
The generating functional of the field strength for arbitrary
two--dimensional orientable and non--orientable surfaces has been calculated
in \cite{16} (and in \cite{17, 18} for the YM$_2$'s). In all of these
calculations, the solutions appear as some infinite summations over the
irreducible representations of the gauge group U($N$) (or SU($N$)). In the
large--$N$ limit, however, these summations are replaced by suitable path
integrals over continuous parameters characterizing the Young tableaux, and
a saddle--point analysis shows that the only significant representation is
the {\it classical} one, which maximizes some effective action.

The continuous function characterizing the representation is a constrained
function, as the length of the rows of the Yang tableau is nonincreasing.
This means that the problem of minimizing the effective action is a
constrained one, and in fact the constraint is a nonholonomic one. For small
values of the surface area, the classical solution satisfies the constraint;
for large values of the surface area, it does not, and the dominating
representation is not the one which minimizes the effective action. This
introduces a phase transition between these two regions, which has been
studied in (for example) \cite{19,20} for YM$_2$. This same problem has been
studied for some special cases of gYM$_2$'s \cite{12}. In \cite{13}, the
density function $\rho$ has been obtained for an arbitrary gYM$_2$ in the
weak--coupling region. $\rho$ is essentially the density function of the
length of the rows of the Young tableau, and by weak--field region it is
meant that the surface area $A$ is less than some critical area
$A_c$. Moreover, calculating $\rho$ in the strong region $A>A_c$ (near the
critical point $A_c$), it is shown that the phase transition of the theory
corresponding to the interaction $\sum_j (n_j-j+N)^4$ is of the third order,
the same behavior seen for YM$_2$. Finally in \cite{14}, the phase
transition of $\sum_j (n_j-j+N)^6$ and $\sum_j (n_j-j+N)^2+g(n_j-j+N)^4$
(for $g<<1$) has been proved to be of the third order. In all of these
works, the procedure of investigating the phase transition is based on the
direct calculation of $\rho_s$ (the density function in the strong region
$A>A_c$). The form of this function is complicated and obtaining a closed
form for it, for arbitrary gYM$_2$'s, is not easy, even near the critical
point.

In this paper we are going to investigate the order of the phase transition
for a typical gYM$_2$, by a different method. In this method, only the
behavior of $\rho_w$ (the density function in the weak--coupling region)
is needed, and in fact this behavior is needed only near the critical point.
In this way, one finds the phase transition of {\it almost} all of gYM$_2$'s
is of the third order. Nontypical gYM$_2$'s are also investigated and the
criteria for a theory to be typical, that is its phase transition be of
the third order are also studied. It is shown that if the second derivative
of the weak--region density function is negative at its absolute maximum
for $A=A_c$, and
if its first derivative with respect to $A$ is positive at the same point,
then the theory exhibits third order phase transition. It is further shown
that theories with monomial interactions satisfy these criteria. It is
also shown that if the second criterion is violated, then the order of
transition will be an integer times 3. Finally, a theory with quadratic
as well as quartic interaction is analyzed and it is shown that it is
typical except for a specific value of the coupling constant, where it
exhibits a transition of the order $5/2$.

Let us investigate this nontypical behavior more closely. It was stated
that if, at the critical point, the first derivative of the absolute
maximum of the density function with respect to $A$ is positive, and the
second derivative of the density function at its absolute maximum is
negative, then the theory is typical, that is the order of the transition is
$3$. What is known, is that at the critical point the absolute maximum of
the density function is an increasing function of $A$. The reason is that
for $A<A_c$, this absolute maximum is less than one, as the system is in
the weak region. For $A>A_c$, this absolute maximum should be greater than
one. So it should pass one at $A=A_c$. This shows that the first derivative
of this absolute maximum is nonnegative, but it may be zero at $A=A_c$.
Theories for which this derivative is zero exactly at $A=A_c$, are
fine tuned. Normally, this derivative is positive. The second criterion for
a theory to be typical can also be discussed the same way. At the absolute
maximum of the density function, the first derivative of the density
function is, of course, zero. The second derivative should be nonpositive.
Theories for which this second derivative is zero exactly at the transition
point, are fine tuned. Other theories are typical and exhibit a third order
phase transition. In fact, in the example which shows a transition of the
order (i.e. exponent) $5/2$, exactly this criterion is violated. The parameters
of
a theory are tuned so that at the critical point, the second derivative of the
density function is zero, and the first nonzero derivative of that function
is the fourth derivative.

The scheme of the paper is the following. In section 2, after a review of
gYM$_2$'s, the criteria for the existence of the third order phase
transition is studied and the order of the transition for nontypical
theories is also studied. In section 3, it is shown that theories with
monomial interaction are typical. In section 4, a theory with a quadratic
and quartic interaction is studied, and it is shown that it exhibits third
order phase transition except for a specific value of the coupling constant.
It is shown that for this value, the order of the transition is $5/2$.
Finally, in section 5 the standard technique is used to investigate the
critical behavior of this theory, and again it is checked that the order of
the transition is $5/2$ at this specific value of the coupling constant.

\section{ Typical gYM$_2$ theories}

The partition function of the gYM$_2$ on a sphere is \cite{10, 16}
\be\label{1}
Z=\sum_r d_r^2e^{-A\Lambda (r)},
\ee
where $r$'s label the irreducible representations of the gauge group, $d_r$
is the dimension of the representation $r$, $A$ is the area of the sphere,
and $\Lambda (r) $ is :
\be\label{2}
\Lambda (r) = \sum_{k=1}^p {a_k \over {N^{k-1}}}C_k(r),
\ee
in which $C_k$ is the $k$'th Casimir of group, and $a_k$'s are arbitrary
constants. Now consider the gauge group U$(N)$ and parameterize its
representation by the integers $n_1\geq n_2\geq\cdots\geq n_N$. It is found
that
\bea\label{3}
d_r&=&\prod_{1\leq i < j \leq N }\left(1+{{n_i-n_j}\over {j-i}}\right),\cr
C_k&=&\sum_{i=1}^N[(n_i+N-i)^k-(N-i)^k].
\eea
For the partition function (1) to be convergent, it is necessary that $p$ in
(2) be even and $a_p$ be positive.

In the  large--$N$ limit, the above summation is replaced by a path
integration over the continuous function
\be\label{4}
\phi (x):=-n(x)-1+x,
\ee
where
\be\label{5}
0 \leq x:=i/N \leq 1 \qquad{\rm and }\qquad n(x):=n_i/N.
\ee
The partition function (\ref{1}) then becomes
\be\label{6}
Z=\int \prod_{0\leq x \leq 1} d\phi(x)e^{S[\phi(x)]},
\ee
where
\be\label{7}
S(\phi)=N^2\left\{-A\int^1_0dxG[\phi(x)]+\int_0^1dx\int_0^1dy\;\log|\phi (x)-
\phi (y)|\right\},
\ee
apart from an unimportant constant, and
\be\label{8}
G(\phi )=\sum_{k=1}^p(-1)^ka_k\phi^k.
\ee
Introducing the density function
\be\label{9}
\rho [\phi (x)]={dx \over d\phi (x)},
\ee
it is seen that it satisfies
\be\label{10}
\int^a_b\rho (z)dz=1,
\ee
where $[b,a]$ is the interval corresponding to the values of $\phi$. If
$G(\phi )$, and therefore $\rho (z)$, is even, then $b=-a$. The condition
$n_1\geq n_2\geq\cdots\geq n_N$ demands
\be\label{11}
\rho (z) \leq 1.
\ee
As $N\rightarrow \infty$, only that representation contributes to the
partition function (\ref{6}) which maximizes $S$. This representation is
found to be specified by the solution of the following integral equation
\cite{13}.
\be\label{12}
g(z)={\rm P}\int^a_b {\rho (z')dz'\over{z-z'}},
\ee
where ${\rm P}$ indicates the principal value of the integral, and
\be\label{13}
g(z):={A\over 2} G'(z).
\ee
The free energy of the theory is defined through
\be\label{14}
F:=-{1\over N^2}{\rm ln}Z,
\ee
from which we have
\be\label{15}
F'(A)=\int_b^adz\; G(z)\rho (z) .
\ee
Using the standard method of solving the integral equation (\ref{12}), the
density function $\rho (z)$ is obtained in terms of the two parameters $a$
and $b$, where $a$ and $b$ themselves must be calculated by solving two
other equations \cite{13}. If $G$ is even, the results are
\be\label{16}
\rho (z)={\sqrt{a^2-z^2}\over \pi}\sum_{n,q=0}^\infty
{(2n-1)!!\over 2^nn!(2n+q+1)!}a^{2n}z^qg^{(2n+q+1)}(0),
\ee
and
\be\label{17}
\sum_{n=0}^\infty {(2n-1)!!\over 2^nn!(2n-1)!}a^{2n}g^{(2n-1)}(0)=1,
\ee
where $g^{(n)}$ is the $n$th derivative of $g(z)$. As it is seen from
(\ref{17}), one cannot find the closed form of $a$ for arbitrary gYM$_2$'s
(that is, for arbitrary $G$'s).

The function $\rho (z)$ found from (\ref{16}), obviously depends on the area
$A$. Its absolute maximum is an increasing function of $A$. But the
constraint (\ref{11}) tells that this maximum must not exceed 1. So, the
function $\rho (z)$ obtained from (\ref{16}) is valid only for $A$ less than
some critical value $A_c$. $A_c$ is the value of $A$ at which the maximum of
$\rho$ becomes 1. The region $A<A_c$ is called the weak--coupling region. In
the region $A>A_c$ (the strong--coupling region), the following ansatz for
$\rho(z)$ is used \cite{13}.
\be\label{18}
\rho_s(z)=\cases{ 1,&$z\in L'$  \cr
\tilde{\rho}_s(z),&$ z\in L,$\cr
}
\ee
where $L'$ is some interval containing the point (or points) maximizing
$\rho$. The form of $\rho_s$ is usually found to consist of some integrals
not expressible in terms of elementary functions. In special cases, it has
been calculated perturbatively for areas slightly more than the critical
area (\cite{19, 20, 13, 14} for example). Having obtained $\rho_s$, the
difference between the free energies in strong and weak regimes ($F_s-F_w$),
for  areas slightly more than the critical area is calculated to obtain the
order of the transition.

Consider this same problem of phase transition from a different point
of view. Denote by $\rho_w(z)$ the weak--region density function,
and by $\rho_s(z)$ the density function (\ref{18}). Suppose that $\rho_w(z)$
attains its (absolute) maximum at $z=z_0$. We have
\be\label{19}
\rho_w(z)=\rho_w(z_0)+{1\over 2}\rho''_w(z_0)(z-z_0)^2+\cdots,
\ee
in which $\rho''_w(z_0)$ is assumed to be nonzero. For $A$ slightly more
than $A_c$, $\rho_w(z_0)$ is slightly more than 1. We want to find
the points $z_1$ satisfying $\rho_w(z_1)=1$. It is easy to see that
\be\label{20}
|z_1-z_0|\approx\sqrt{2[1-\rho_w(z_0)]/\rho''_w(z_0)}\sim\alpha^{1/2},
\ee
where $\alpha:=\rho_w(z_0)-1$. Next, consider $F_s-F_w$:
\bea\label{21}
F[\rho_s]-F[\rho_w]&=&\int\left[ {{\delta F}\over{\delta\rho (z)}}
\right]_{\rho_w}[\rho_s(z)-\rho_w(z)]dz\cr
&&+{1\over 2}\int\left[ {{\delta^2 F}\over{\delta\rho (z)\delta\rho (z')}}
\right]_{\rho_w}[\rho_s(z)-\rho_w(z)]\cr
&&\times[\rho_s(z')-\rho_w(z')]dz\; dz'+\cdots .
\eea
As $\rho_w$ is the classical solution, it satisfies
\be
\left({{\delta S}\over{\delta\rho}}\right)_{\rho_w}=\lambda ,
\ee
where $\lambda$ is the Lagrange--multiplier used to introduce the condition
(\ref{10}) in the action $S$. So the first term of the right--hand side of
(\ref{21}) becomes
\bea\label{22}
\int\left[{{\delta F}\over{\delta\rho (z)}}\right]_{\rho_w}[\rho_s(z)-
\rho_w(z)]dz&=&-{\lambda\over{N^2}}\left[\int\rho_s(z)dz-\int\rho_w(z)dz
\right]\cr
&=&0,
\eea
in which we have used the fact that $\rho_w$ and $\rho_s$, both satisfy
(\ref{10}) ---perhaps with different integration regions, but it is not
important, as one may define $\rho$ to be zero outside the integration
region and then extend the integration region to the whole real line. Next,
consider the second term of (\ref{21}). One can divide the integration
region into three parts:\\
$R_1$: $z,z'\in L_s$,\\
$R_2$: $z,z'\in L_w$,\\
$R_3$: $z\in L_s$ and $z'\in L_w$ or visa versa,\\
where $L_s$ is the region in which $\rho_w>1$, and $L_w$ is the remaining
region. We have
\be\label{101}
\rho_w(z)-\rho_s(z)\sim\alpha,\qquad z\in L_s.
\ee
From this, it is seen that
\be\label{102}
\int_{L_w}[\rho_w(z)-\rho_s(z)]dz\sim\alpha^{3/2},
\ee
as the width of the region $L_s$ is of the order $\alpha^{1/2}$ (\ref{20}).
Using (\ref{10}) for $\rho_s$ and $\rho_w$, it is seen that
\be\label{23}
\int_{L_w}[\rho_s(z)-\rho_w(z)]dz\sim\alpha^{3/2}.
\ee
The width of $L_w$ is of the order $\alpha^0$. This shows that on the
average,
\be\label{103}
\rho_s(z)-\rho_w(z)\sim\alpha^{3/2},\qquad z\in L_w.
\ee
Using these, it is seen that in $R_1$, the integrand in the second term of
(\ref{21}) is of the order $\alpha^2$, while the area of the integration
region is of the order $\alpha$. So the integral in this region is of the
order $\alpha^3$. In $R_2$, the integrand is of the order
$\alpha\times\alpha^{3/2}$, and the area of $R_2$ is of the order
$\alpha^{1/2}$. So the integral in this region too is of the order
$\alpha^3$. Finally, in $R_3$, the integrand is of the order
$\alpha^{3/2}\times\alpha^{3/2}$, while th area of $R_3$ is of the order
$\alpha^0$. So, the integral in this region is of the order $\alpha^3$ as
well. One therefor deduces
\be\label{24}
F_s-F_w\sim\alpha^3.
\ee
Here we have used $\rho''_w(z_0)\neq 0$.

To find the critical behavior of the system, we must obtain $F_s-F_w$ as a
function of $A-A_c$. To do so, consider $\alpha$ as a function of $A$ and
expand it around $A=A_c$:
\be\label{25}
\alpha (A)=\alpha(A_c)+\alpha'(A_c)(A-A_c)+\cdots .
\ee
But $\alpha (A_c)=0$. So if $\alpha'(A_c)\neq 0$, we have
\be\label{104}
\alpha\sim (A-A_c),
\ee
and from this
\be\label{26}
F_s-F_w\sim (A-A_c)^3,
\ee
that is, the model has a third--order phase transition. What we have proved,
is that if $\rho''_w(z_0)\neq 0$ and $\alpha'(A_c)\neq 0$, then the order of
the phase transition is 3. In fact, to prove this we have made an implicit
assumption that (\ref{103}) holds not only on the average, but at {\it
almost} every point in $L_w$. There might be very special cases where the
difference between $\rho_s$ and $\rho_w$ is larger than $\alpha^3$ but changes
sign exactly in a such a manner that the integral of this difference is of
the order $\alpha^{3/2}$. Also note that the whole reasoning is independent
of the number of points at them $\rho_w$ attains its absolute maximum.

Continuing to make this assumption, let's see what happens if either
$\rho''_w(z_0)$ or $\alpha'(A_c)$ is zero. As $z_0$ is the point which
maximizes $\rho_w$, the first nonzero derivative of $\rho_w$ at this point
should be an even derivative. Let it be the $(2m)$-th derivative. Also,
$\alpha$ should be increasing with respect to $A$ at $A_c$. This shows that
its first nonzero derivative should be on odd derivative. Let it be the
$(2l-1)$-th derivative. It follows that the width of $L_s$ is of the order
$\alpha^{1/(2m)}$, and from this
\be\label{27}
F_s-F_w\sim\alpha^{2+(1/m)}.
\ee
One also has
\be\label{105}
\alpha\sim (A-A_c)^{2l-1}.
\ee
Combining these two, one arrives at
\be\label{106}
F_s-F_w\sim (A-A_c)^{(2l-1)[2+(1/m)]},
\ee
where $m$ and $l$ are positive integers.

\section {$G(\phi )=\phi^{2k}$ gYM$_2$'s}

For $G(\phi )=\phi^{2k}$ ($k$ an integer greater than 1), it has been shown
that the density function $\rho_w(z)$, (\ref{16}), has a minimum at $z=0$
and two symmetric maxima at points $z_0=\pm a_k \sqrt{y_k}$, \cite{14}.
$a_k$ is the parameter appearing in (\ref{16}), and can be evaluated from
(\ref{17}):
\be\label{28}
a_k=\left[{{2^k(k-1)!}\over{A(2k-1)!!}}\right]^{{1\over{2k}}}.
\ee
$y_k$ is independent of $a_k$ and is determined from
\be\label{29}
-1+\sum_{n=0}^{k-2}{{(2n-1)!!}\over{2^{n+1}(n+1)!}}y_k^{-(n+1)}=0.
\ee
Now it is possible to find the $A$--dependence of $\rho (z_0)$. Using
(\ref{16}),
\be\label{30}
\rho_w (z)={{kA}\over{\pi}}{\sqrt{a^2_k-z^2}}\sum_{n=0}^{k-1}
{{(2n-1)!!}\over{2^nn!}}a_k^{2n}z^{2k-2n-2}.
\ee
From this,
\bea\label{31}
\rho_w(z_0)&=&A^{{1\over 2k}}{{k\sqrt{1-y_k}}\over\pi}
\left[{{2^k(k-1)!}\over{(2k-1)!!}}\right]^{{2k-1}\over{2k}}
\sum_{n=0}^{k-1}{{(2n-1)!!}\over{2^nn!}}y_k^{k-n-1}\cr
&=:&A^{{1\over 2k}}f(k),
\eea
where $f(k)$ is independent of $A$. $A_c$ is determined by $\rho_k(z_0)=1$,
which yields
\be\label{32}
A_c={1\over{[f(k)]^{2k}}}.
\ee
From this,
\be\label{107}
\alpha (A)=\left({A\over{A_c}}\right)^{1\over{2k}}-1,
\ee
which shows that the derivative of $\alpha$ with respect to $A$ never
vanishes. It is also easy to check that the $\rho''_w(z_0)$ doesn't vanish
either. Using $\rho'_w(z_0)=0$, it is seen that
\bea\label{33}
\rho''_w(z_0)&=&{{kAa^{2k-3}}\over\pi}{{(2k-1)u}\over{\sqrt{1-u^2}}}
\left[ -(2k-1)u^{2k-3}+\sum_{n=0}^{k-2}{{(2n-1)!!}\over{2^{n+1}(n+1)!}}
u^{2k-2n-5}\right]\cr
&=&-{{kAa^{2k-3}}\over\pi}{{2k-1}\over{\sqrt{1-u^2}}}
\sum_{n=0}^{k-2}{{(2n-1)!!}\over{2^n n!}}u^{2k-2n-4},
\eea
which is clearly negative. Here
\be\label{108}
u:={{z_0}\over a},
\ee
and (\ref{29}) has been used. We conclude all gYM$_2$'s with
$G(\phi )=\phi^{2k}$ gYM$_2$'s exhibit third--order phase transition.

\section {The model $G(\phi )=\phi^2+g\phi^4$}

We now consider $G(\phi )=\phi^2+g\phi^4$, with $g>0$ so that in this $G$
has a minimum. Using (\ref{16}), we have
\be\label{34}
\rho_w (z)={A\over \pi}{\sqrt{a^2-z^2}}(1+ga^2+2gz^2),
\ee
and from (\ref{17}),
\be\label{35}
{1\over 2}Aa^2+{3\over 4}gAa^4=1.
\ee
The shape of $\rho_w (z)$ depends on $A$. For $3ga^2<1$, or $A>4g$ using
(\ref{35}), $\rho_w(z)$ has only one maximum at $z=0$. For $A<4g$, it has a
minimum at $z=0$ and two maxima at $\pm z_0$. We consider three distinct
cases.
\subsection{$g<A_c/4$}
In this case, $\rho_w$ has two maxima for small areas. However, if the area
exceeds $4g<A_c$, $\rho_w$ will have only one maximum. As this happens
before the critical area, the phase transition properties are governed by a
one--maximum density function. The maximum of this function occurs at
$z=0$. Evaluating $a$ from (\ref{35}), we arrive at
\be\label{36}
\rho_w(0)={A\over{2\pi}}\left( 2+\sqrt{1+12{g\over A}}\right)
\sqrt{{3\over g}\left( -1+\sqrt{1+12{g\over A}}\right)},
\ee
and from this
\be\label{37}
\alpha'(A_c)={1\over\sqrt{27\pi^2 g}}\; {{(3g/A_c)-1+\sqrt{1+12g/A_c}}\over
{\sqrt{-1+\sqrt{1+12g/A_c}}}}>0.
\ee
It is also seen that
\be\label{120}
\rho''_w(0)={A\over{\pi a}}(3ga^2-1)<0.
\ee
So the order of the transition is 3.
\subsection{$g>A_c/4$}
In this case, for areas up to $4g>A_c$, the density function has two
maxima at $\pm z_0$:
\be\label{38}
z_0=\sqrt{{1\over{6g}}\left( -2+\sqrt{1+12{g\over A}}\right)},
\ee
From this, one obtains $\rho_w(z_0)$, and then
\be\label{39}
\alpha'(A_c)=\sqrt{2\over{27g\pi^2}} \; {{1+3g/A_c}\over
{(1+12g/A_c)^{1/4}}}>0.
\ee
It is also not difficult to check that $\rho''_w(z_0)$ is negative. So in
this case too, the order of the phase--transition is 3.
\subsection{$g=A_c/4$}
In this case, for $A<4g=A_c$ the function $\rho_w(z)$ has two maxima at
$\pm z_0$ and one minimum at $z=0$. As $A$ is increased, the difference
$\rho_w(z_0)-\rho_w(0)$ becomes smaller, and at $A=A_c$ it becomes zero.
For $A>A_c$, $\rho_w(z)$ has only one maximum at $z=0$. At $A=A_c$, the
maximum of the density function becomes 1 (and occurs at $z=0$). Using
this, the form of $\rho_w$ from (\ref{34}), and (\ref{35}), it is seen that
\be\label{41}
A_c={{27}\over{64}}\pi^2,
\ee
\be\label{42}
g_0={A_c\over 4}={{27}\over{256}}\pi^2,
\ee
and
\be\label{43}
a_c={{16}\over{9\pi}}.
\ee
The interesting point for $g=g_0$ is that $\rho''_w(0)|_{A=A_c}=0$. In
fact, one can see that
\bea\label{109}
\rho''_w(0)&=&{{3g_0Aa(a+a_c)}\over\pi}(a-a_c)\cr
&=&-{4\over{3\pi}}(A-A_c)+\cdots ,
\eea
while $\rho^{(4)}_w(0)$ is negative. So for $z_1$, satisfying
$\rho_w(z_1)=1$, or
\be\label{45}
{1\over{2!}}\rho''_w(0)z_1^2+{1\over{4!}}\rho^{(4)}_w(0)z_1^4=1-\rho_w(0),
\ee
one obtains
\be\label{110}
|z_1|\sim\alpha^{1/4}.
\ee
One also has
\be\label{111}
\alpha ={{28}\over{27\pi^2}}(A-A_c)+\cdots ,
\ee
from which it is seen that
\be\label{112}
|z_1|\sim (A-A_c)^{1/4}.
\ee
This is an example of the case $m=2$ and $l=1$, discussed at the end of
section 2. It follows then that
\be\label{48}
F_s-F_w\sim (A-A_c)^{5/2}.
\ee
The model $G(\phi )=\phi^2+(27\pi^2/64)\phi^4$ exhibits thus a phase
transition of order 5/2.

\section {Explicit evaluation of the order of the phase transition of
the model $G(\phi )=\phi^2+{27\pi^2\over 64}\phi^4$}

We now try to study the phase transition of the model
$G(\phi )=\phi^2+g_0\phi^4$ explicitly ($g_0=27\pi^2/64$). Using (\ref{15})
and (\ref{34}), one finds that in the weak region
\be\label{49}
F_w'(A)={1\over 8}a^4A+{5\over{16}}g_0a^6A+{9\over{64}}g_0^2a^8A,
\ee
where $a$ can be found from (\ref{35}). In the strong region, one uses the
following ansatz for $\rho_s(z)$
\be\label{50}
\rho_s(z)=\cases{ 1,&$z\in [-b,b] $  \cr
\tilde{\rho}_s(z),&$ z\in [-a,-b]\bigcup [b,a], $\cr
}
\ee
(as $\rho_w(z)$ has only one maximum for $A>A_c$). Using the standard method
of solving the integral equations obtained (see \cite{14} for more details)
one arrives at
\bea\label{51}
F_s'(A)&=&\left({1\over 8}M^2-{1\over 2}N\right)\int_{-b}^b
{{d\lambda}\over{U(\lambda)}}+{1\over 2}M\int_{-b}^b
{{\lambda^2 d\lambda}\over{U(\lambda)}}-\int_{-b}^b
{{\lambda^4 d\lambda}\over{U(\lambda)}}\cr
&& +g_0 \left[\left(-{1\over 4}MN+{1\over{16}}M^3\right)\int_{-b}^b
{{d\lambda}\over{U(\lambda)}}\right.\cr
&&+\left({1\over 8}M^2 -{1\over 2}N\right)\int_{-b}^b
{{\lambda^2 d\lambda}\over{U(\lambda)}}\cr
&&\left. +{1\over 2}M\int_{-b}^b{{\lambda^4 d\lambda}\over{U(\lambda)}}
-\int_{-b}^b{{\lambda^6 d\lambda}\over{U(\lambda)}} \right]\cr
&&-2g_0A\left({1\over 4}MN-{1\over{16}}M^3\right)\cr
&&-2g_0^2A\left( -{1\over 8}N^2+{3\over{16}}M^2N-{5\over{128}}M^4\right),
\eea
in which
\bea\label{52}
U(\lambda)&:=&\sqrt{(a^2-\lambda^2)(b^2-\lambda^2)},\cr
M&:=&a^2+b^2,\cr
N&:=&a^2b^2.
\eea
The parameters $a$ and $b$ are determined from following relations.
\be\label{53}
A+g_0AM-\int_{-b}^b{{d\lambda}\over{U(\lambda)}}=0,
\ee
and
\be\label{54}
{1\over 2}MA+g_0A\left({3\over 4}M^2-N\right) -\int_{-b}^b
{{\lambda^2 d\lambda}\over{U(\lambda)}}=1.
\ee
As we are interested in phase transition, it is sufficient to look at the
solutions for $A$ slightly greater than $A_c$, which we obtain
perturbatively.

To do so, we use the following change of variables
\be\label{55}
a=:a_c(1-h),
\ee
and
\be\label{56}
(b/a)^2=:v,
\ee
in which $a_c$ is given by (\ref{43}). Eliminating $A$ from (\ref{53}) and
(\ref{54}), and solving the result for $h$, one obtains
\be\label{57}
h={{27}\over{224}}v^2+{{131}\over{6272}}v^3+\cdots,
\ee
from this, and (\ref{53}), one obtains $A$ in terms of $v$:
\be\label{113}
{{A-A_c}\over{A_c}}={9\over{28}}v^2+{{109}\over{784}}v^3+\cdots.
\ee
Note that $A-A_c$ is proportional to $b^4$ (in the leading order), as
expected. Substituting these in (\ref{51}), one obtains $F'_s$:
\be\label{58}
F_s'={1\over{35721\pi^2}}(36848-10836v^2+6289v^3+\cdots ).
\ee
From (\ref{35}), one obtains $a$ in terms of $A$. Substituting this in
(\ref{49}), and using (\ref{113}) as a {\it definition} for $v$, one obtains
\be\label{59}
F_w'={1\over{35721\pi^2}}(36848-10836v^2-4687v^3+\cdots ),
\ee
from which
\be\label{60}
F_s'(A)-F_w'(A)={{224}\over{729\pi^2}}v^3+\cdots.
\ee
Using (\ref{113}), one can write this as
\be\label{63}
F_s'(A)-F_w'(A)={{196\sqrt{7}}\over{729A_c}}\left({{A-A_c}\over{A_c}}\right)
^{3/2}+\cdots .
\ee
which proves that the order of the phase transition is 5/2, in agreement
with (\ref{48}).
\vspace {1cm}

\noindent{\bf Acknowledgement}

\noindent M. Alimohammadi would like to thank the research council of the
University of Tehran, for partial financial support. The authors would
also like to thank Institute for Studies in Theoretical Physics and
Mathematics for partial financial support.

\end{document}